\documentstyle[12pt]{article}
\begin{document}

\begin{center}

{\large \bf DENSITY MATRIX KINETIC EQUATION DESCRIBING A
PASSAGE OF FAST ATOMIC SYSTEMS THROUGH MATTER}\\
\vspace{1.25cm}

O Voskresenskaya
\footnote{Current address: Institut f\"ur
Theoretische Physik, Universit\"at Heidelberg, Max-Plank-Institut
f\"ur Kernphysik, D - 69029 Heidelberg, Deutschland;
e-mail: Olga.Voskresenskaja@mpi-hd.mpg.de}
\vspace{.25cm}

{\small Joint Institute for Nuclear Research, Dubna, Moscow
Region, 141980 Russia}\\
\vspace{.75cm}

\begin{minipage}{13cm}
{\small \bf Abstract.}
{\small The quantum-mechanical consideration of
a passage of fast dimesoatoms through matter is given. A set of
quantum-kinetic equations for the density matrix elements describing
their internal state evolution is derived. It is shown that
probabilistic description of internal dynamics of hydrogen-like
atoms is impossible even at sufficiently low energies because
of the ``accidental'' degeneracy of their energy levels.}
\end{minipage}

\end{center}

\vspace{.25cm}

\section{Introduction}

In the last years interest in studying elementary hadronic atoms
(hydrogen-like atoms formed by two oppositely charged hadrons)
has been growing \cite{HadAtom}.

The basic properties of these atoms are governed by the Coulomb
interaction between their constituents. The strong hadron-hadron
interaction causes only shifts of energy levels and increase of
widths of $s$-states of hadronic atoms which are related in a
simple way to the values of strong scattering lengths of hadrons.

Thus, the study of hadronic atoms provides valuable information
about the low-energy hadronic interactions.
At present a number of ongoing or planned experiments are aimed
to measure strong energy level shifts and (or) widths of the ground
states of the atoms  $A_{\pi^- p}$, $A_{K^- p}$, $A_{\pi^-\pi^+}$,
$A_{\pi^{\pm}K^{\mp}}$ with a high precision \cite{PSI,KEK,dirac}.

The  $A_{\pi^-p}$ and $A_{K^{-}p}$ atoms can be produced at rest
in the laboratory frame, and a well-developed technique of X-ray
spectroscopy can be applied for a precise measurement of their
radiative spectra.

The study of the atoms  $A_{\pi^-\pi^+}$ and $A_{\pi^{\pm}K^{\mp}}$
(dimesoatoms) is a much more complica\-te problem because they
practically cannot be produced at rest. The only available way
to produce a reasonable amount of dimesoatoms is the interaction of
high-energy projectiles with fixed nuclear targets. The dimesoatoms
produced in this way move with high (relativistic, as a rule)
velocities which determines the choice of methods for investigation
of their properties.

For example, the method for measurement of $\tau_{0}$ (the pionium
ground state lifetime) used in the DIRAC experiment at CERN
\cite{dirac} is based on comparison of the observed rate of
$\pi^+\pi^-$-pairs from the break-up of pionium atoms in the Coulomb
field of target atoms with the theoretical calculations of this rate
which include the $\tau_{0}$ as a parameter.

It is clear that the error in the value of $\tau_{0}$ resulting
from such indirect measurement includes not only experimental errors
but also uncertainties of the approximations used in the theoretical
description of the pionium production and their subsequent passage
through the target matter.

Thus, any improvement of the theoretical treatment of the problems
related to the DIRAC experiment helps to reduce the resulting error
of $\tau_{0}$ measurement.  Some important results in  this direction
have been obtained in \cite{AVY,traut,MROW,VOSK01} dealing with the
theoretical description of the production of dimesoatoms and their
interacti\-ons with separate target atoms.

In this paper, we will touch on a problem of describing  the internal
dynamics of dimesoatoms  moving through matter and interacting with
its atoms. It is commonly supposed \cite{AT} that the dimesoatomic
dynamics can be described with the help of a set of kinetic equations
for the probabilities $P_i(z)$ to find the dimesotom in the definite
quantum state $|i\rangle$ at the distance $z$ from the production
point.

Although the form of these equations is similar to the Pauli quantum
kinetic equations \cite{pauli}, we will refer to them as the classical
(probabilistic) ones because they ignore interference between different
states of dimesoatoms during their passage through the matter which is
pure quantum effects.

It is well known that these interference effects play a decisive
role in giving rise to the so-called superpenetration phenomena
\cite{nemen81} at ultrarelativistic energies. However, at the typical
dimesoatom energies in DIRAC most interference effects are strongly
suppressed by the time-formation effects.

But in the case of dimesoatoms some of interference effects occur
even at sufficiently low energies because of accidental degeneracy
of energy levels of hydrogen-like atoms \cite{land}. Since these
effects cannot be treated in the classical approach of the paper
\cite{AT}, one have to use a more sophisticated approach for
consideration of internal dimesoatom dynamics based on the
density-matrix description of quantum systems.

The main goal of this paper is to derive a set of equations
for the density matrix elements describing the internal dynamics
of multilevel atoms moving through matter and interacting with its
atoms.  Although for simplicity the derivation is restricted to
the case of elementary (two-particle) atoms, the method
can be easily generalized to a more complicated case of arbitrary
composed system.

The plan of the paper is as follows. In Section 2, the main
approximations are described.  In Section 3, a set of equations for
density matrix elements is derived.  In Section 4, the correspondence
between density-matrix approach and probabilistic approach of the paper
\cite{AT} is discussed. In Section 5, a brief summary of the main
results is given.

\section{Main approximations }

Any discussion of relativistic dimesotom interactions faces, first
of all, the problem of the relativistic description of a bound system.
Application of usual methods based on the Bethe-Salpeter equation
\cite{salpet} would lead to overcomplication of the technical part of
the problem under consideration, which may mask the main goal of this
work.

To avoid this difficulty, we will use the following trick. We will
first consider the interaction of fast but still nonrelativistic
dimesoatoms with the target. In this case the relatively simple
mathematical formalism of nonrelativistic quantum mechanics can be
used for the description of both internal and external dynamics of
dimesoatoms. This results in simple energy dependence of the density
matrix.

The next step is the ``analytical'' extension of this nonrelativistic
result to relativistic energies appealing to the physical intuition.
Although this way of derivation cannot be considered as rigorous  we
believe that it allows one to get the most essential features of the
desired result.

We restrict our consideration to the case of rather high dimesoatom
velocities and rather thin targets so that the ``frozen'' target
approximation can be used. In other words, we will suppose that the
positions of target atoms practically do not change during the passage
of a dimesoatom trough the target.

In this case, the stationary Schr\"odinger equation can be used to
describe the interaction between the dimesoatom and the target:
\begin{equation}
\label{eq:d1}
H\psi=E\psi\,,
\end{equation}
\begin{equation}
\label{eq:d2}
H=H_{int}+H_{ext}\,,
\end{equation}
\begin{equation}
\label{eq:d3}
H_{int}=-\frac{1}{2\mu}\Delta_{\vec r}+ V_{int}(\vec r)\,,
\end{equation}
\begin{equation}
\label{eq:d4}
H_{ext}=-\frac{1}{2M}\Delta_{\vec R}+ V_{ext}(\vec r,\vec R;
\{\vec\rho\})\,.
\end{equation}
Here $\psi\equiv\psi(\vec r_+,\vec r_-)$,
$\vec r_+$ and $\vec r_-$ are the radius-vectors of the positively
and negatively charged mesons respectively, $m_+$ and $m_-$ are their
masses; $M=m_++m_-$,  $\mu=m_+m_-/M$; $\vec R=(m_+\vec r_++m_-\vec r_-)/
M$ is the radius-vector of the dimesoatom center-of-mass; $\vec r=(\vec
r_+-\vec r_-)$ is the radius-vector of relative motion of dimesoatom
constituents; $\{\vec\rho\}$ is the set of the radius-vectors of the
target atom center-of-mass; $V_{int}(\vec r)$ is the potential energy
of interaction between dimesoatom constituents; $V_{ext}(\vec r,\vec R;
\{\vec\rho\})$ is the potential energy of interaction between the
dimesoatom and the target atom:
\begin{equation}
\label{eq:d6}
V_{ext}=e\sum_k[\Phi(\vec
r_+-\vec\rho_k)-\Phi(\vec r_--\vec\rho_k)]\,,
\end{equation}
\begin{equation}
\label{eq:d7}
\vec r_+=\vec R+\xi\vec r,\quad \vec r_-=\vec R-\eta\vec r\,,
\end{equation}
\begin{equation}
\label{eq:d8}
\xi=\frac{m_-}{M},\quad \eta=\frac{m_+}{M}=1-\xi\,,
\end{equation}
$\Phi$ is the electrostatic potential of the target atom at rest;
$e$ is the elementary charge.

Let us consider that the beam of dimesoatoms is in a definite state
of the discrete spectrum $|i\rangle$ and moves in the positive
direction of the $Oz$-axis ($\vec p=(0,0,p)$ is the beam momentum).
The target is supposed to be a plate of thickness $L$ and of the
infinite transverse size, placed perpendicular to the beam, so that
one surface has $z=0$ and the other has $z=L$. Then, it the $z\leq 0$,
\begin{equation}
\label{eq:d13}
\psi(\vec r_+,\vec r_-)=\exp(\vec p\vec R)\cdot \phi_i(\vec r)\,,
\end{equation}
\begin{equation}
\label{eq:d14}
H_{int}\phi_i(\vec r)=\varepsilon_i\phi_i(\vec r)\,,
\end{equation}
where $\varepsilon_i$ is the energy of the state $|i\rangle$.

The solution of (\ref{eq:d1}) in the region $z>0$ will be sought
in the form
\begin{equation}
\label{eq:d15}
\psi(\vec r_+,\vec r_-)=\exp(i\vec p\vec R)\cdot
F(\vec R,\vec r,\{\vec\rho\})\,.
\end{equation}

Then the equation for $F$ following from (\ref{eq:d1})
is conveniently written as
\begin{eqnarray}
\label{eq:d16}
&&\frac{1}{2M}
\left(\frac{\partial^2}{\partial z^2}+ 2ip
\frac{\partial}{\partial z}\right)F\left(z,\vec b,\vec r,\{\vec
\rho\}\right)\nonumber\\
&=&\left(-\frac{\Delta_{\vec b}}{2M}+
H_{int}+V_{ext}\right)
F\left(z,\vec b,\vec r,\{\vec\rho\}\right)\,,
\end{eqnarray}
where $z$ and $\vec b$ are the longitudinal and transverse parts
of the radius-vector $\vec R=\{\vec b,z\}$.

At rather high dimesoatomic energies, such that the condition
\begin{equation}
pa\gg 1,\quad a\sim(m_e\cdot\alpha\cdot Z^{1/3})^{-1}
\label{eq:d17}
\end{equation}
is satisfied, the term $\partial^2F/\partial z^2$ on the left side
of (\ref{eq:d16}) can be neglected \cite{land} (in (\ref{eq:d17})
$a$ is the screening radius of the target atom, $Z$ is its atomic
number and $m_e$ is the electron mass).

In this approximation the quantity
\begin{equation}
\tilde F=\exp\left(\frac{iM}{p}\varepsilon_iz\right)\cdot F
\label{eq:d18}
\end{equation}
obeys the equation
\begin{equation}
\label{eq:d19}
-i\frac{\partial\tilde F}{\partial z}= \tilde H\cdot\tilde F\,,
\end{equation}
\begin{equation}
\label{eq:20}
\tilde H=-\frac{1}{2p}\Delta_{\vec b}+\frac{M}{p}H_{int}+\frac{1}{v}
V_{ext}
\end{equation}
which looks like the nonstationary Schr\"odinger equation.

In (\ref {eq:20}) $v$ is the dimesoatom velocity, which in the
nonrelativistic limit coinsides with the ratio $p/M$.  But keeping in
mind the subsequent ``analytical extension'' of the below results to
the region of relativistic energies, we will assume not the equality
$p/M=v$ in the general case but the relativistic relationship
$p=M v\gamma$. With this prescription the right energy dependence
of the time-formation effects is reproduced (see also \cite{voskr}).

\section{Density matrix kinetic equation}

Let us write the expression for the density matrix describing
the dimesoatom state after covering the distance $z$ in the matter
\begin{eqnarray}
\rho(z;\vec r_1,\vec r_2)&=&\Sigma^{-1}\int
d\vec b d\vec b_1^{\,\prime} d\vec b_2^{\,\prime}d\vec r_1^{\,\prime}
d\vec r_2^{\,\prime}\left\langle
G(z;\vec b,\vec r_1;\vec b_1^{\,\prime},\vec r_1^{\,\prime};
  \{\vec\rho\})\right.\nonumber\\
&&\left.\times G^{\ast}(z;\vec b,\vec r_2;\vec b_2^{\,\prime},\vec
r_2^{\,\prime};\{\vec\rho\})\right\rangle_{\{\vec\rho\}}
\phi_i(\vec r_1^{\,\prime})\phi^{\ast}_i(\vec r_2^{\,\prime})\,.
\label{eq:d35}
\end{eqnarray}
Here $G$ is the Green function of (\ref{eq:d19}), the solution of
which is given by the expression
\begin{equation}
\tilde F(z,\vec b,\vec r,\{\vec\rho\})=\int  G(z;\vec b,\vec r;
\vec b^{\,\prime},\vec r^{\,\prime};\{\vec\rho\})\phi_i
(\vec r^{\,\prime}) d\vec r^{\,\prime}d\vec b^{\,\prime}\,,
\label{eq:d22}
\end{equation}
$\Sigma$ is the beam cross section area, and the symbol
$\langle\rangle_{\{\vec \rho\}}$ means averaging over all possible
states of target atoms.

To perform this averaging in the explicit form, we will use
the path integral representation \cite{fein} of the Green
function:
\begin{equation}
G(z;\vec b,\vec r;\vec b^{\,\prime},\vec
r^{\,\prime};\{\vec\rho\,\}) = \int D\vec b(\tilde z)D\vec r(\tilde
z)\exp(iS)\,,
\label{eq:d23}
\end{equation}
where the action $S$ of the dimesoatom interacting with the matter
is
\begin{eqnarray} S&=&\int\limits_{0}^{z}dz^{\,\prime}
\left[ \frac{p}{2}\left(\frac{d\vec b(z^{\,\prime})}{dz^{\,\prime}}
\right)^2  +\frac{v\gamma\mu}{2}
\left(\frac{d\vec r(z^{\,\prime})}{dz^{\,\prime}}\right)^2\right.\nonumber\\
&&\left.-\frac{1}{v\gamma}V_{int}\left(\vec r(z^{\,\prime})\right)
-\frac{1}{v}V_{ext}
\left(z^{\,\prime};\vec b(z^{\,\prime}),\vec r(z^{\,\prime})\right)
\right]\,.
\label{eq:d24}
\end{eqnarray}
The path integration in (\ref{eq:d23}) is performed over all
trajectories $\left\{\vec b(\tilde z),\vec r(\tilde z) \right\}$
beginning with the point $\left\{\vec b(0),\vec r(0) \right\}=
\left\{\vec b^{\,\prime},\vec r^{\,\prime}\right\}$ and ending with the
point $\left\{\vec b(z),\vec r(z) \right\}=\left\{\vec b,\vec r\right\}$.

It was shown in \cite{VOSKR99} that
\begin{eqnarray}
\int\limits_{0}^{z}dz^{\,\prime}\frac{1}{v}V_{ext}
\left(z^{\,\prime};\vec b(z^{\,\prime}),\vec r(z^{\,\prime})\right)
&=&\sum_k\vartheta(z-z_k)\left\{\chi\left(\vec b(z_k)+\xi\vec s(z_k)-
\vec\tau_k \right)\right.\nonumber\\
&&\left.-\chi\left(\vec b(z_k)-\eta\vec s(z_k)-\vec\tau_k
\right)\right\}\,,
\label{eq:d25}
\end{eqnarray}
where the Heavyside step-function $\vartheta(z)$ is $0$ for $z<0$ and
$1$ for $z>0$; $\vec s$  and $\vec\tau_k$  are the transverse
components of the vectors $\vec r$ and $\vec\rho_k$; $z_k$ is the
longitudinal component of the vector $\vec\rho_k$.

The quantity
\begin{eqnarray}
\chi(\vec B_{\xi(\eta)})=\frac{e}{v}\int\limits_{-\infty}^{\infty}\Phi
\left(\sqrt{\vec B_{\xi(\eta)}^2+z^2}\right)dz\,,\\
\vec B_{\xi}=\vec b+\xi\vec s-\vec\tau_k,\quad
\vec B_{\eta}=\vec b-\eta\vec s-\vec\tau_k\,, \nonumber
\label{eq:d26}
\end{eqnarray}
is the phase shift acquired by the wave function of the hadron
while it moves through the field of the isolated target atom.

Let us use the relation  \cite{nemen81}
\begin{equation}
\left\langle\exp\left[\sum_k f(\vec\rho_k,\{x\})\right]
\right\rangle_{\{\vec\rho\}}
=\exp\left\{-n_0\int d\vec\rho\left(1-f(\vec\rho,\{x\})\right)
\right\}\,,
\label{eq:d27}
\end{equation}
where $n_0$ is the number of target atoms in the unite volume;
$f(\vec\rho_k,\{x\})$  is the arbitrary function of the target atom
coordinates and of the set $\{x\}$  of others variables.

Then the average of the product of two Green functions over the
target atom coordinates is
\begin{eqnarray}
&&\left\langle G(z;\vec
b_1,\vec r_1;\vec b_1^{\,\prime},\vec r_1^{\,\prime}; \{\vec\rho\})
G^{\ast}(z;\vec b_2,\vec r_2;\vec b_2^{\,\prime},\vec
r_2^{\,\prime};\{\vec\rho\})\right\rangle_{\{\vec\rho\}}\nonumber\\
&&=\int D\vec b_1(\tilde z)D\vec b_2(\tilde z)
D\vec r_1(\tilde z)D\vec r_1(\tilde z)\exp\left(i\bar S_1-i\bar S_2-
n_0\int\limits_{0}^{z}dz^{\,\prime}\int\omega d\vec\tau\right),
\label{eq:d28}
\end{eqnarray}
\begin{equation}
\bar S_{1(2)}=\int\limits_{0}^{z}dz^{\,\prime}
\left[
\frac{p}{2}\left(\frac{d\vec b_{1(2)}(z^{\,\prime})}{dz^{\,\prime}}\right)^2
+\frac{v\gamma\mu}{2}\left(
 \frac{d\vec r_{1(2)}(z^{\,\prime})}{dz^{\,\prime}}\right)^2
-\frac{1}{v\gamma}V_{int}
  \left(\vec r_{1(2)}(z^{\,\prime}) \right)
\right],
\label{eq:d29}
\end{equation}
\begin{equation}
\omega=1-\exp[i\Delta\chi_{1}-i\Delta\chi_{2}]\,,
\label{eq:d30}
\end{equation}
\begin{eqnarray}
\Delta\chi_{1(2)}&=&
\chi\left[\vec b_{1(2)}(z^{\,\prime})+\xi\vec s_{1(2)}(z^{\,\prime})
    -\vec\tau\right]
-\chi\left[\vec b_{1(2)}(z^{\,\prime})-\eta\vec s_{1(2)}(z^{\,\prime})
    -\vec\tau\right].
\label{eq:d31}
\end{eqnarray}

With the replacement
\begin{equation}
\vec\tau\to\vec\tau+\vec b_{+}(z^{\,\prime})\,,
\label{eq:d32}
\end{equation}
where the $\vec\tau$ is impact vector and
\begin{equation}
\vec b_{+}(z^{\,\prime})=
\left(\vec b_{1}(z^{\,\prime})+\vec b_{2}(z^{\,\prime})-
\vec\tau\right)\,,
\label{eq:d33}
\end{equation}
it is easy to see that $\omega$ depends only on
\begin{equation}
\vec b_{-}(z^{\,\prime})=
\left(\vec b_{1}(z^{\,\prime})-\vec
b_{2}(z^{\,\prime})-\vec\tau\right)
\label{eq:d34}
\end{equation}
and does not depend on $\vec b_{+}(z^{\,\prime})$.

This allows to perform path integration over the variables
$\vec b_{1}(\tilde z)$ and $\vec b_{2}(\tilde z)$ in (\ref{eq:d28})
in the explicit form (see e.g. \cite{VOSKR99}) and to represent
the density matrix (\ref{eq:d35}) in the form
\begin{eqnarray}
\rho(z;\vec r_1,\vec r_2)&=&\int D\vec r_1(\tilde z)D\vec r_2(\tilde z)
d\vec r_1^{\,\prime}d\vec r_2^{\,\prime}\nonumber\\
&&\times\exp\left\{i\bar S_1-i\bar S_2-
n_0\int\limits_{0}^{z}dz^{\,\prime}\widetilde{\Omega}\right\}
\phi_i(\vec r_1^{\;\prime})\phi^{\ast}_i(\vec r_2^{\;\prime})\,,
\label{eq:d36}
\end{eqnarray}
where
\begin{equation}
\widetilde{\Omega}=\int \omega d\tau=
\Omega\bigl(\vec s_{1}(z^{\,\prime}),\vec s_{2}(z^{\,\prime})\bigl)\,,
\label{eq:d37}
\end{equation}
\begin{eqnarray}
\Omega(\vec s_{1},\vec s_{2})&=&
\int d\tau\left\{
\Gamma(\vec\tau,\vec s_1)+\Gamma^{\ast}(\vec\tau,\vec s_2)
-\Gamma(\vec\tau,\vec s_1)\Gamma^{\ast}(\vec\tau,\vec s_2)
\right\}\,,
\label{eq:d38}
\end{eqnarray}
\begin{eqnarray}
\Gamma(\vec\tau,\vec s_{1(2)})&=&
1-\exp\left\{i\chi(\vec\tau-\xi\vec s_{1(2)})
-i\chi(\vec\tau+\eta\vec s_{1(2)})\right\}\,.
\label{eq:d39}
\end{eqnarray}

$\Gamma(\vec\tau,\vec s)$ is the interaction operator of the Glauber
theory for interaction of dimesoatoms with target atoms \cite{glaub}.
In particular, the transition amplitudes of the dimesoatoms between
the states $|i\rangle$  and $|k\rangle$ in the Coulomb field of target
atoms are related to $\Gamma(\vec\tau,\vec s)$ by the equation
\begin{eqnarray}
A_{ki}(\vec q)&=&\frac{i}{2\pi}\int d^2\tau\,d^3r
e^{i\vec q\vec \tau}\psi_k^{\ast}(\vec r)\psi_i(\vec r)
\Gamma(\vec\tau,\vec s)\,.
\label{eq:d40}
\end{eqnarray}

Path integrations (like the usual ones) in (\ref{eq:d36}) can be
performed by numerical methods (\cite{comput}).

But for the purpose of the qualitative analysis it is more useful to
deal with the kinetic equation for the density matrix
\begin{eqnarray}
i\frac{\partial\rho(z;\vec r_1,\vec r_2)}{\partial z}&=&
\frac{1}{v\gamma}\left[H_{int}(\vec r_1)-
H^{\ast}_{int}(\vec r_2)\right]\rho(z;\vec r_1,\vec r_2)\nonumber\\
&&-in_0 \Omega(\vec s_1,\vec s_2)\rho(z;\vec r_1,\vec r_2)\,,
\label{eq:d41}
\end{eqnarray}
which is the consequence of the path integral representation
(\ref{eq:d36})  (see, for example, \cite{fein}).

Because of the hermiticity of $H_{int}$ this equation can be
rewritten in the operator form
\begin{eqnarray}
i\frac{\partial\rho}{\partial z}=
\frac{1}{v\gamma}\left[H_{int},\rho\right] - in_0 \Omega\rho\,,
\label{eq:d42}
\end{eqnarray}
where the first term on the right side (Liouville term) describes the
causal part of the internal dimesoatom dynamics.  The nature of the
second term is pure stochastic because of the stochastic distribution
of the atoms in the target.\footnote{ For each separate projectile
the target ``is frozen'' in the definite state. But because of the
stochastic distribution of the projectiles in the beam these definite
``frozen'' states of the target are distributed stochastically too.}

In the dimesoatom rest frame Eq. (\ref{eq:d41}) takes the form
\cite{voskr}
\begin{eqnarray} i\frac{\partial\rho(t;\vec r_1,\vec
r_2)}{\partial t}&=& H_{int}(\vec r_1)\rho(t;\vec r_1,\vec r_2)
-H^{\ast}_{int}(\vec r_2)\rho(t;\vec r_1,\vec r_2)\nonumber\\
&&-iv\gamma n_0 \Omega(\vec s_1,\vec s_2)\rho(t;\vec r_1,\vec r_2)
\label{eq:d43}
\end{eqnarray}
which is similar to the form of Eq.~(116) of paper \cite{chang} for
the density matrix of atoms moving in the laser fields.  But the
meaning of the terms on the right side of this equation and on the
right side of our Eq. (\ref{eq:d43}) is different.

The causal part of (116) also includes the term describing
interaction of atoms with the laser fields while the stochastic term
of this equation describes effects of spontaneous relaxation caused
by interaction of atoms with the quantized electromagnetic field.

Strictly speaking, a similar term should be added to the right
part of (\ref{eq:d43}) too. But it can be shown that its influence
on the internal dynamics of the dimesoatoms is negligible compared
with similar effects caused by interaction of dimesoatoms with
target atoms.\footnote{Interaction of dimesoatoms with the quantized
electromagnetic field causes only deexcitation of excited dimesoatoms
through emission of real photons. Interaction of dimesoatoms with
target atoms causes both excitation and deexcitation of dimesoatoms
through exchange of virtual photons with target atoms.}
For this reason we do not include the effects of
spontaneous electromagnetic relaxation in our consideration.

There is effect that play important role in the dimesoatom
dynamics which is not taken into account in (\ref{eq:d43}).
It is the effect of instability of dimesoatoms caused by the
possibility of their annihilation into neutral hadrons.
We will include it phenomenologically.

With this aim let us consider the density-matrix elements
\begin{equation}
\label{eq:4.46}
\rho_{ik}(z)=\int\psi_i^{\ast}(\vec r)\psi_k(\vec r^{\,\prime})
\rho(z;\vec r,\vec r^{\,\prime})d\vec rd\vec r^{\,\prime}\,.
\end{equation}

Taking into account the Schr\"odinger equation
\begin{equation}
\label{eq:4.47}
H_{int}\psi_i=\varepsilon_i\psi_i
\end{equation}
it is easy to see that matrix Eq.~(\ref{eq:d41}) is equivalent
to the following system of equations for the density-matrix elements:
\begin{equation}
\label{eq:4.48}
\frac{\partial\rho_{ik}}{\partial z}=i\Delta_{ik}\rho_{ik}
-n_0\sum_{l,m}\Omega_{ik,lm}\rho_{lm}\,,
\end{equation}
where
$$\Delta_{ik}=(\varepsilon_k-\varepsilon_i)/v\gamma,$$
\begin{equation}
\label{eq:4.49}
\Omega_{ik,lm}=\int\psi_i^{\ast}(\vec r)\psi_l(\vec r)
\psi_k(\vec r^{\,\prime})\psi^{\ast}_m(\vec r^{\,\prime})
\Omega(\vec s,\vec s^{\,\prime})d\vec rd\vec r^{\;\prime}\,.
\end{equation}

It is evident that after including the instability effects
the solution of this system at $n_0=0$ (``empty'' target)
\begin{eqnarray}
\label{eq:4.50}
\rho_{ik}(z)&=&\rho_{ik}(0)\exp[i\Delta_{ik}\cdot z]
\end{eqnarray}
is modified to
\begin{eqnarray}
\label{eq:4.51}
\rho_{ik}(z)&=&\rho_{ik}(0)\exp
\left\{\frac{z}{v\gamma}\left[i(\varepsilon_{k}-\varepsilon_{i})
-\frac{1}{2}(\Gamma_i+\Gamma_k)\right]\right\}\,,
\end{eqnarray}
where $\Gamma_{i(k)}$ are the widths of the states $|i(k)\rangle$.

Hence Eq. (\ref{eq:4.48}) must be replaced by
\begin{eqnarray}
\label{eq:4.53}
\frac{\partial\rho_{ik}(z)}{\partial z}&=&\frac{1}{v\gamma}
\left[i(\varepsilon_{k}-\varepsilon_{i})-\frac{1}{2}
(\Gamma_i+\Gamma_k)\right]
\rho_{ik}(z)-n_0\sum_{l,m}\Omega_{ik,lm}\rho_{lm}(z)\,.
\end{eqnarray}

For the subsequent discussion it is convenient to represent
(\ref{eq:4.53}) in the integral form
\begin{equation}
\label{eq:4.54}
\rho_{ik}=\rho_{ik}(0)\exp[-a_{ik}\cdot z]
-n_0\sum_{l,m}\Omega_{ik,lm}\int\exp[-a_{ik}\cdot (z-z^{\prime})]
\rho_{lm}(z^{\prime})dz^{\prime} \,,
\end{equation}
\begin{equation}
\label{eq:d55}
a_{ik}=-\frac{1}{v\gamma}\left[i(\varepsilon_{k}-\varepsilon_{i})
-\frac{1}{2}(\Gamma_i+\Gamma_k)\right]\,.
\end{equation}

\section{Time-formation effects}

It is easy to check that the coefficients $\Omega_{ik,lm}$ (\ref{eq:4.49})
are different from zero if the relations
\begin{equation} \label{eq:4.55}
m_i-m_k-m_l+m_m=0\,,\quad l_i-l_k-l_l+l_m=2s
\end{equation}
are satisfied. Above, $m_{i(k,l,m)}$ and $l_{i(k,l,m)}$
are the magnetic and orbital quantum numbers of the states
$|i(k,l,m)\rangle$ and $s$ is an arbitrary integral number (the quntization
axis is chosen to coinside with the $Oz$-axis).

Since the dimesoatoms are produced practically only in the $ns$-states
($l=0,m=0$) \cite{dirac}, the  ``selection rules''  for the quantities
$\rho_{ik}(z)$
\begin{equation}
\label{eq:4.56}
\rho_{ik}(z)\ne 0\quad\mbox{if and only if}\quad m_i=m_k\,,
\quad l_i=l_k+2s
\end{equation}
follow from (\ref{eq:4.50}), (\ref{eq:4.55}).

No restrictions on the values $n_{i(k)}$ --- principal quantum
numbers of the states $|i(k)\rangle$ --- are imposed by the dynamics
of interaction of dimesoatoms with target atoms represented by the
quantities $\Omega_{ik,lm}$.  But they are imposed by the so-called
``length (time)-formation effects'',\footnote{These effects belong
to the wide class of the effects systematic investigation of which
began in the papers \cite{TM,LP} (further development see in
\cite{tamm}).} represented by the oscillating factors
$\exp[i\Delta_{ik}\cdot (z-z^{\prime})]$ in (\ref{eq:4.54}).

At super-high energies, such that $\Delta_{ik}L\ll 1$ ($L$ is
the thickness of the target), the factors $\exp[i\Delta_{ik}\cdot
(z-z^{\prime})]\sim 1$ for any values of the indices $i,k$,  and
the relations between the values of the diagonal and off-diagonal
density matrix elements $\rho_{ik}(z)$ are governed only by the
properties of the coefficients $\Omega_{ik,lm}$.

For ``low'' energies (such that $\Delta_{ik}\bar \lambda\gg 1$,
where $\bar \lambda$ is the typical dimesoatom free-path length
in matter) the numerical values of the off-diagonal density matrix
elements $\rho_{ik}(z)$ ($i\ne k$) become negligible compared to
the diagonal ones if $\varepsilon_i\ne\varepsilon_k$.

If energy levels of the system under consideration are non-degenerate
($\varepsilon_i\ne \varepsilon_k$, if $i\ne k$), only the diagonal
density matrix elements $\rho_{ii}=P_{i}$ survive at ``low'' energies
because of length-formation effects, and the  internal dynamics of such
systems can be described by a system of kinetic equations for
probabilities $P_i(z)$
\begin{equation}
\label{eq:4.57}
\frac{dP_i}{dz}=\sum_{i,l}c_{il}P_{l}(z),
\quad c_{il}=-n_0\Omega_{ii,ll}-a_{ii}\delta_{il}
\end{equation}
which coinside with the system equations of \cite{AT}.

Taking into account the ``selection rules'' (\ref{eq:4.56}),
we clearly see that this statement is also valid when there is only
degeneracy with respect to the value of the magnetic quantum number
that is, if the hadron-hadron interaction has only the central
symmetry.

However, the Coulomb interaction in the hydrogen-like atoms has
additional symmetry that causes extra (accidental) degeneracy
of the energy levels with respect to the value of the orbital
quantum number. Because of this, even at very low energies
some off-diagonal density matrix elements $\rho_{ik}(z)$ with
\begin{equation}
\label{eq:4.59}
\varepsilon_i= \varepsilon_k,\quad l_i=l_k+2s
\end{equation}
survive together with diagonal ones.
This excludes the possibility of describing of internal dynamics
of these atoms in the framework of ``classical'' (probabilistic)
approach \cite{AT} at any energies.\footnote{Strictly speaking,
strong interaction between dimesoatomic constituents eliminates
the accidental degeneracy of the dimesoatomic energy levels.
But this effects is so small numerically that it has no
significance for the problem under consideration (for example,
($\varepsilon_{2p}-\varepsilon_{2s})/\varepsilon_{2s}\sim 10^{-3}$).}

\section{Conclusion}

The accurate description of internal dynamics of relativistic
dimesoatoms moving through matter requires the density matrix
formalism.
The kinetic equation for the density matrix could be represented
in the form of the Liouville equation only if the positions of the
target atoms will change in time according to a definite law.
Because of the stochastic change of these positions, the part of the
kinetic equation for density matrix describing  the interaction of
dimesoatoms with target atoms becomes more complicate.
The main result of the present paper is the derivation in the
explicit form (\ref{eq:d38}) of the term describing this interaction.
That fact that we have automatically obtained an expression for
(\ref{eq:d38}) in terms of interaction operators of the Glauber theory
can be considered as additional confirmation of the validity of the
approximations used.

\section*{Acknowledgments}

The author is grateful to Leonid Nemenov for stimulating interest
to the work, and to Alexander Tarasov for fruitful discussions and
invaluable help in preparation of the manuscript for publication.
I would like also to express my thank to Leonid Afanasyev for 
careful reading the manuscript and useful remarks and corrections.
I am acknowledged to the Institute for Theoretical Physics at Heidelberg
University and the MPI f\"ur Kernphysik, Heidelberg, where some essential
part of this work  was done, for support. Also, it is a pleasure for me
to thank  J\"org H\"ufner, Boris Kopeliovich and J\"org Raufeisen for
their hospitality in Heidelberg and interesting discussions.

\end{document}